\documentclass[11pt,aps,showpacs,preprintnumbers,amsmath,amssymb,nofootinbib,preprint]{article}

\usepackage{epsfig}
\usepackage{dcolumn}
\usepackage{amsmath}
\usepackage{color}
\usepackage{graphicx}

\newcommand{\ba} {\begin{eqnarray}}
\newcommand{\ea} {\end{eqnarray}}
\def \be  {\begin{equation}}
\def \ee  {\end{equation}}
\def \ee  {\end{equation}}
\def \bea {\begin{eqnarray}}
\def \eea {\end{eqnarray}}

\renewcommand{\figurename}{{\bf Fig.}}
\renewcommand{\tablename}{{\bf Tab.}}
\usepackage{graphicx}
\usepackage{dcolumn} 
\usepackage{bm}      
\begin{document}
\renewcommand{\figurename}{Fig.}
\renewcommand{\tablename}{Tab.}
\title{\vspace*{-1.cm}\hfill {\small\bf ECTP-2015-17} \\ \hfill {\small\bf WLCAPP-2015-17}\\
\vspace*{2cm}
QCD Phase-transition and chemical freezeout in nonzero magnetic field at NICA}

\author{Abdel Nasser~Tawfik\footnote{Corresponding author: a.tawfik@eng.mti.edu.eg}  \\
{ \it Egyptian Center for Theoretical Physics (ECTP),}\\
{\it Modern University for Technology and Information (MTI), 11571 Cairo, Egypt} and\\
{\it World Laboratory for Cosmology And Particle Physics (WLCAPP), 11571 Cairo, Egypt. }  
}

\date{\today}
\maketitle

\begin{abstract}
Because of relativistic off-center motion of the charged spectators and the local momentum-imbalance experienced by the participants, a huge magnetic field is likely generated in high-energy collisions. The influence of such short-lived magnetic field  on the QCD phase-transition(s) shall be analysed. From Polyakov linear-sigma model, we study the chiral phase-transition and the magnetic response and susceptibility in dependence on temperature, density and magnetic field strength. The systematic measurements of the phase-transition characterizing signals, such as the fluctuations, the dynamical correlations and the in-medium modifications of rho-meson, for instance, in different interacting systems and collision centralities are conjectured to reveal an almost complete description for the QCD phase-structure and the chemical freezeout. We limit the discussion to NICA energies.
\end{abstract}

                           \hspace*{2.62cm} ~Viscous properties of QGP\\ 
{\bf \hspace*{0.45cm} PACS Nos:}~11.10.Wx,25.75.Nq,98.62.En,12.38.Cy

\maketitle

\section{Introduction}

In most of the experimental signals revealing the deconfinement phase-transition such as the correlations and the elliptic flow of produced particles, the increase in the collision size and centrality is conjectured to improve their detection. The heavy-ion collisions - on the other hand - introduce a remarkable effect; a nonzero magnetic field, which can be generated due to the relativistic off-center motion of the spectators electric charges and due to the local imbalance in the momentum carried by colliding nucleons, which leads to local angular-momentum \cite{prephiral2,prephiral3}. Accordingly, the life span of the resulting magnetic field can approximatively be estimated from the duration of both mechanisms. The relativistic off-center motion of the electric charges and the local momentum imbalance of colliding nuclei keep up with the strong interaction causing them, i.e., $\sim 10^{-23}~$sec. Thus, the resulting magnetic field is short-lived but its strength can be very huge, ${\cal O}(m_{\pi}^2)$ \cite{prephiral1}. We focus the discussion to NICA energies, where $e B \simeq 0.156\, m_{\pi}^2\simeq 1.567 \times 10^{17}~$Gauss, Fig. \ref{fig:eBdepence}. 

Quantum electrodynamical (QED) phenomena such as magnetic catalysis which describes how the magnetic field generates masses, dynamically and the Meissner effect in which the magnetic field changes the order of the phase transition in type-I superconductor are borrowed from solid-state physics in order to study the possible influence of the generated magnetic field on the quantum choromodynamical (QCD) phase-space structure \cite{fraga} and/or the response of the hadronic and partonic matter to nonzero magnetic field in thermal and dense medium. 

We have utilized the Polyakov linear-sigma model (PLSM) in characterizing various QCD phenomena. Thermodynamic quantities including the higher-order moments of particle multiplicities are studied and compared with lattice QCD results \cite{Tawfik:quasi,Tawfik:2014uka}. The chiral phase-structure of (pseudo)-scalar and (axial)-vector meson masses in thermal and dense medium was analysed, as well \cite{Tawfik:2014gga}. Furthermore, we have reported on some properties of the quark-gluon plasma (QGP) in nonzero magnetic field \cite{Tawfik:2014hwa}. From these studies, various parameters of PLSM can be fixed. Based on this PLSM can also be utilized at high density, where the non-perturbative lattice technique becomes no longer applicable.

In the present work, we shall describe how the chiral phase-transition and the magnetic response and susceptibility are affected at high density and nonzero magnetic field. In doing this, we utilize PLSM in the mean field approximation. Upon availability, our calculations shall be compared with lattice QCD simulations.

\section{A short reminder to Polyakov linear-sigma model in finite magnetic filed}

By assuming that the magnetic field ($e B$) is directed along $z$-axis of the collision plane and from magnetic catalysis and Landau quantization \cite{Shovkovy2013}, the magnetic field strength causes a considerable modification to the quark (and equivalently in antiquark) dispersion relations, 
\begin{eqnarray}
E_{B,f} &=& [p_{z}^{2}+m_{f}^{2}+2 |q_{f}|(n-S_z+1/2) B]^{1/2}, \label{eq:moddisp}
\end{eqnarray}
where subscript $f$ runs over all quarks (and antiquarks) flavors, $S_z$ is the component of the spin in the magnetic field direction, $n$ being index label for Landau levels and $|q_f|>0$ is the electric charge of $f$-th quark flavor. 
From the magnetic catalysis property, a dimensional reduction can be implemented \cite{Shovkovy2013}, 
\begin{eqnarray}
\int \frac{d^3 p}{(2 \pi)^3} &\longrightarrow& \frac{|q_{f}| B}{2 \pi} \sum_{\nu=0}^{\infty} \int \frac{d p_z}{2 \pi} (2- \delta_{0\nu}),
\end{eqnarray} 
where $2-\delta_{0\nu}$ stands for degenerate Landau levels and $n-S_z+1/2$ is replaced by the quantum number $\nu$. 

Further details about the PLSM formalism can be taken from Ref. \cite{Tawfik:quasi,Tawfik:2014uka,Tawfik:2014gga}. 
In the mean field approximation, the PLSM grand potential can be summarized as \cite{Tawfik:2014uka},
\begin{eqnarray}
\Omega(T, \mu) &=& U(\sigma_x, \sigma_y) + \mathbf{\mathcal{U}}(\phi, \phi^*, T) + \Omega_{\bar{\psi}
\psi} (T;\phi,\phi^{*},B). \hspace*{6mm} \label{eq:plsmPOT}
\end{eqnarray} 
The mesonic potential $\mathbf{\mathcal{U}}(\phi, \phi^*, T)$ is assumed to vanish at high temperatures $\geq \Lambda_{QCD}$. The purely mesonic potential is given as \cite{Schaefer:2008hk}. 
 \begin{eqnarray}
U(\sigma_x, \sigma_y) &=& - h_l \sigma_x - h_s \sigma_y + \frac{m^2}{2}\, (\sigma^2_x+\sigma^2_y) - \frac{c}{2\sqrt{2}} \sigma^2_x \sigma_y  
+ \frac{\lambda_1}{2} \, \sigma^2_x \sigma^2_y +\frac{(2 \lambda_1 +\lambda_2)}{8} \sigma^4_x  + \frac{(\lambda_1+\lambda_2)}{4}\sigma^4_y. \hspace*{8mm} \label{Upotio}
\label{pure:meson}
\end{eqnarray}

Due to the Landau quantization and the magnetic catalysis, the quarks and antiquark contributions to the PLSM potential are given as  \cite{Tawfik:2014hwa} 
\begin{eqnarray}
\Omega_{\bar{q}q}(T, \mu _f, B) &=& - 2 \sum_{f=l, s} \frac{|q_f| B \, T}{(2 \pi)^2} \,  \sum_{\nu = 0}^{\nu _{max_{f}}}  (2-\delta _{0 \nu })    \int_0^{\infty} dp_z \nonumber \\ & & 
\left\{ \ln \left[ 1+3\left(\phi+\phi^* e^{F(x)}\right)\; e^{-F(x)} +e^{-3 F(x)}\right]  +
\right. \nonumber \\ &&  \left.
\hspace*{2mm}\ln \left[ 1+3\left(\phi^*+\phi e^{-F(-x)}\right)\; e^{-F(-x)}+e^{-F(-x)}\right] \right\}, \hspace*{4mm} \label{new-qqpotio}
\end{eqnarray}
where $F(x)=-(E_{B, f} -\mu _f)/T$ and $F(-x)=-(E_{B, f} +\mu _f)/T$. $E_{B,f}$ is given in Eq. (\ref{eq:moddisp}) and $\mu_f$ refers to the chemical potential of $f$-th quark flavor. In vanishing magnetic field, the quarks potential reads
\begin{eqnarray} 
\Omega_{\bar{q}q}(T, \mu _f) &=& -2 \,T \sum_{f=l, s} \int_0^{\infty} \frac{d^3\vec{p}}{(2 \pi)^3} \nonumber \\
&& \left\{ \ln \left[ 1+3\left(\phi+\phi^* e^{-F(x)}\right)\times e^{-F(x)}+e^{-3 F(x)}\right] +
\right. \nonumber \\ & &  \left. 
\hspace*{2mm} \ln \left[ 1+3\left(\phi^*+\phi e^{-F(-x)}\right)\times e^{-F(-x)}+e^{-3 F(-x)}\right] \right\}, \hspace*{4mm} \label{thermalOMG}
\end{eqnarray}

Landau quantization, which is applied in order to add restrictions to the quarks due to the existence of free charges in the plasma phase, affects the strongly interacting matter by reducing the electromagnetic interactions between the quarks. In nonzero magnetic field, one still has to sum over color and flavor degrees of freedom with a dispersion relation given in Eq. (\ref{eq:moddisp}), because the electric charges of the quarks are not equal, Eq. (\ref{new-qqpotio}). On the other hand, the zero modes, Eq. (\ref{thermalOMG}), are taken into consideration in the present calculations. Furthermore, we have estimated the occupation of Landau levels at different temperatures, densities (chemical potentials), and magnetic field strengths (not shown here). We have concluded that the population of Landau levels is most sensitive to the magnetic field and the quark charges, as well. Having PLSM potential, Eq. (\ref{eq:plsmPOT}), we can then straightforwardly deduce the thermodynamics quantities of interest.

\section{Results and discussion}
\label{sec:RD}

\subsection{Chiral phase-transition at high density and nonzero magnetic field}

\begin{figure}[!htb]
\centering{
\includegraphics[width=7cm]{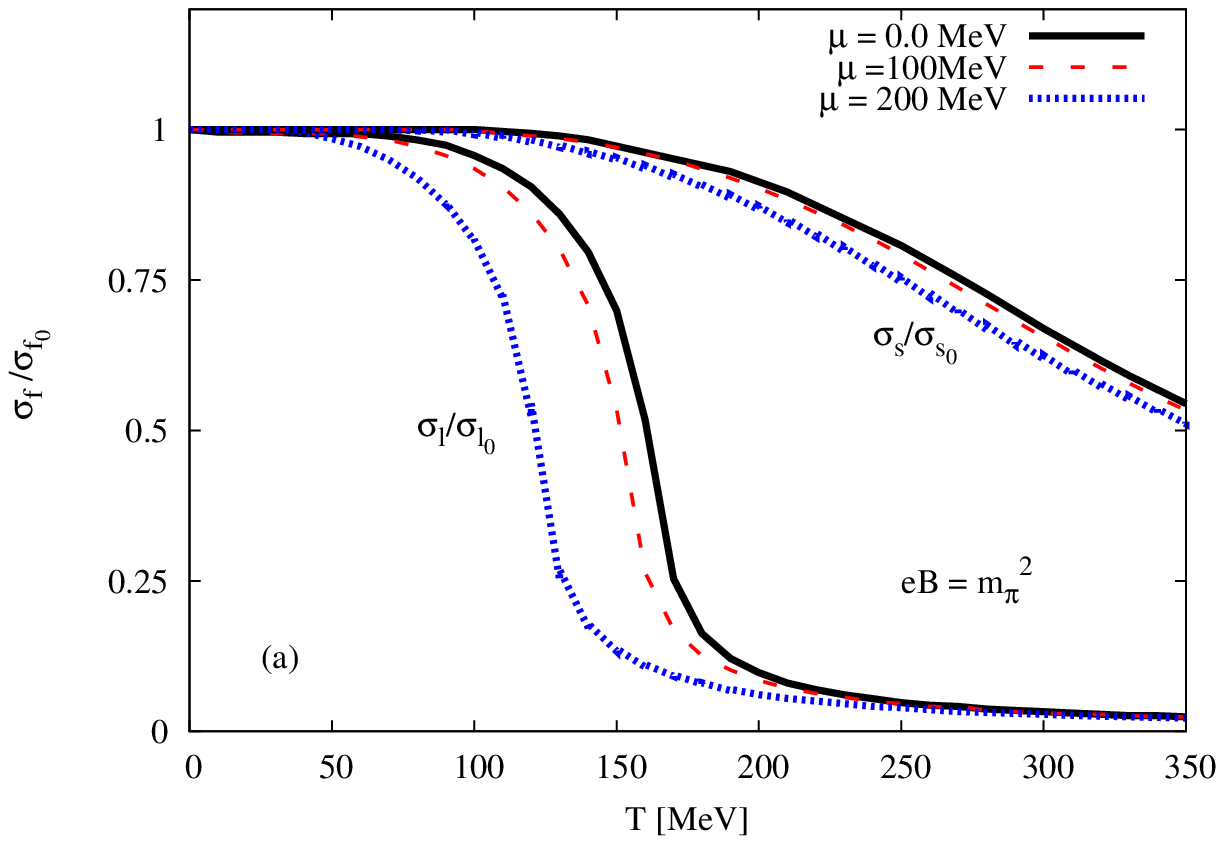}
\includegraphics[width=7cm]{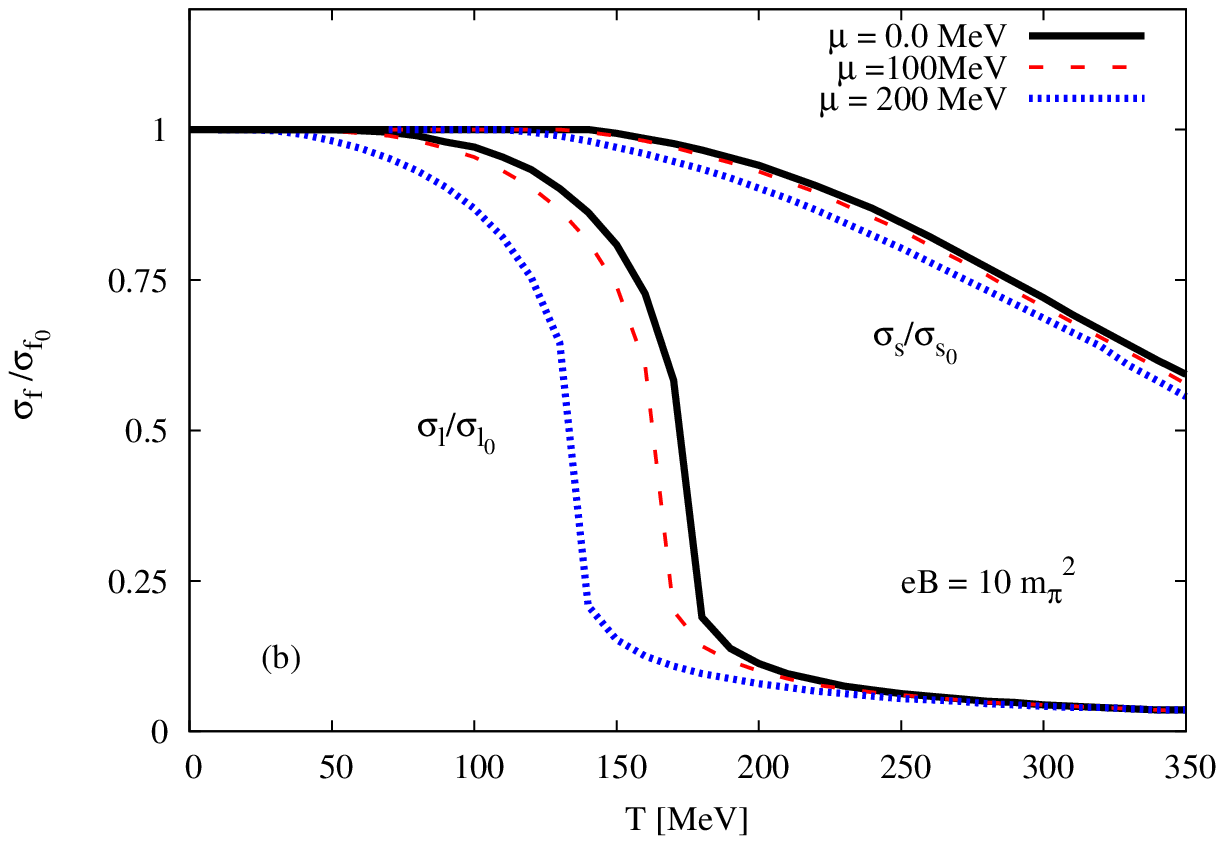}
\caption{The normalized chiral condensate as a function of temperature at $eB=m^2_{\pi}$ (left-hand panel) and $eB=10\, m^2_{\pi}$ (right-hand panel) at $\mu_b=0$ (solid), $100$ (dashed), $200~$MeV (dotted curves).  \label{fig:sbtrc1}  }}
\end{figure}

In Fig. \ref{fig:sbtrc1}, the normalized chiral condensates ($\sigma_{l}/\sigma_{l_o}$ and $\sigma _{s}/\sigma _{s_o}$) are depicted in dependence on the temperature $T$ at different baryon chemical potentials; $\mu_b=0$ (solid), $100$ (dashed), $200~$MeV (dotted curves), and various magnetic field strengths; $eB=m^2_{\pi}$ (left-hand panel) and  $eB=10\, m^2_{\pi}$ (right-hand panel).  In these calculations, the PLSM parameters are determined at (vacuum) sigma-meson mass $800~$MeV, where the  vacuum light  and strange chiral condensates are measured as $\sigma_{l_o}=92.5~$ MeV and $\sigma_{s_o}=94.2~$MeV with $\sigma_{l}\equiv \sigma_x$ and $\sigma_{s}\equiv \sigma_y$, respectively. We notice that the values of the chiral condensates decline (having smaller critical temperatures) with increasing the baryon chemical potential. This means that, the chiral critical temperature decreases with the increase in the magnetic field strength ($e B$). It is obvious that the increasing in the baryon chemical potential ($\mu_b$) has the same effect. In other words, we conclude that the influences of the magnetic field are almost the same as that of the baryon chemical potentials, especially on decreasing the phase-transition temperature.  The influence of the corresponding magnetic field ($e B \simeq 0.156\, m_{\pi}^2\simeq 1.567 \times 10^{17}~$Gauss) becomes stronger with increasing density, e.g. NICA, which operates at the highest baryon density.

\begin{figure}[!htb]
\centering{
\includegraphics[width=7cm]{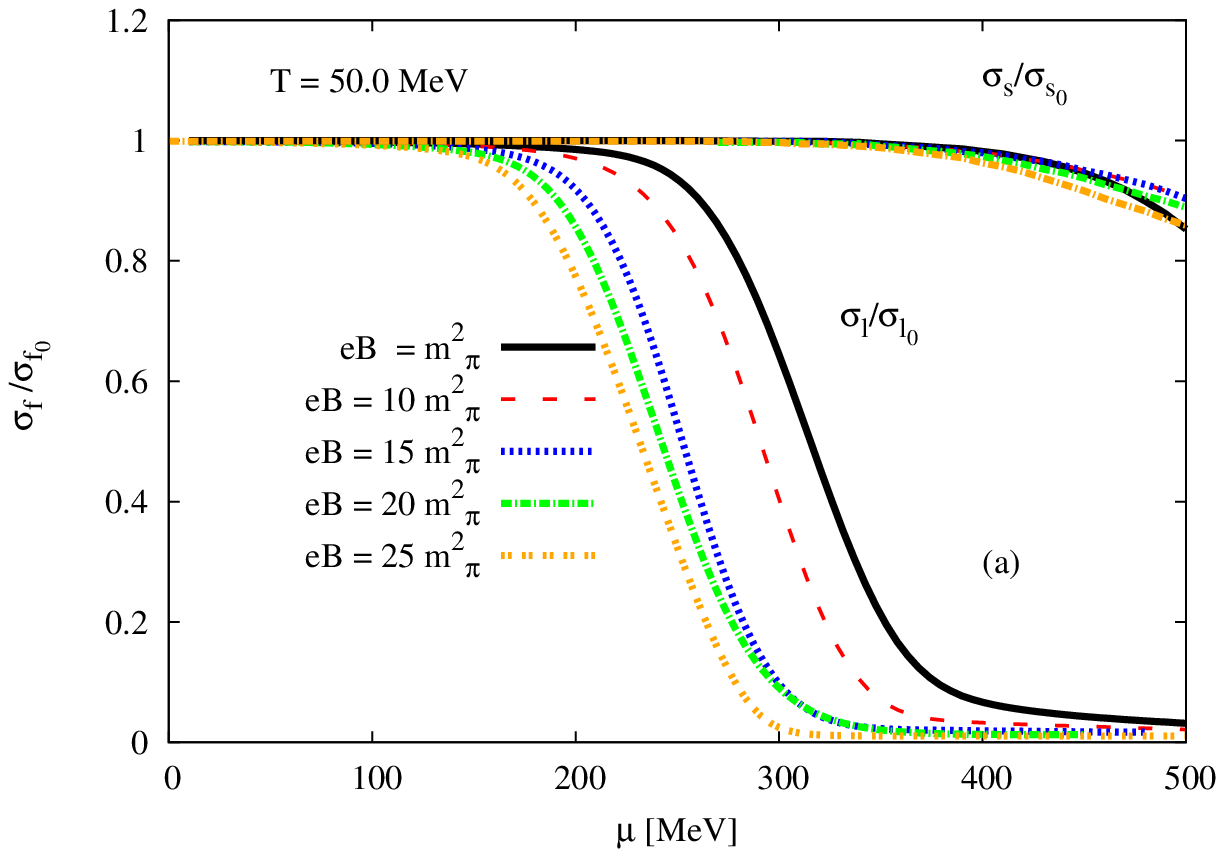}
\includegraphics[width=7cm]{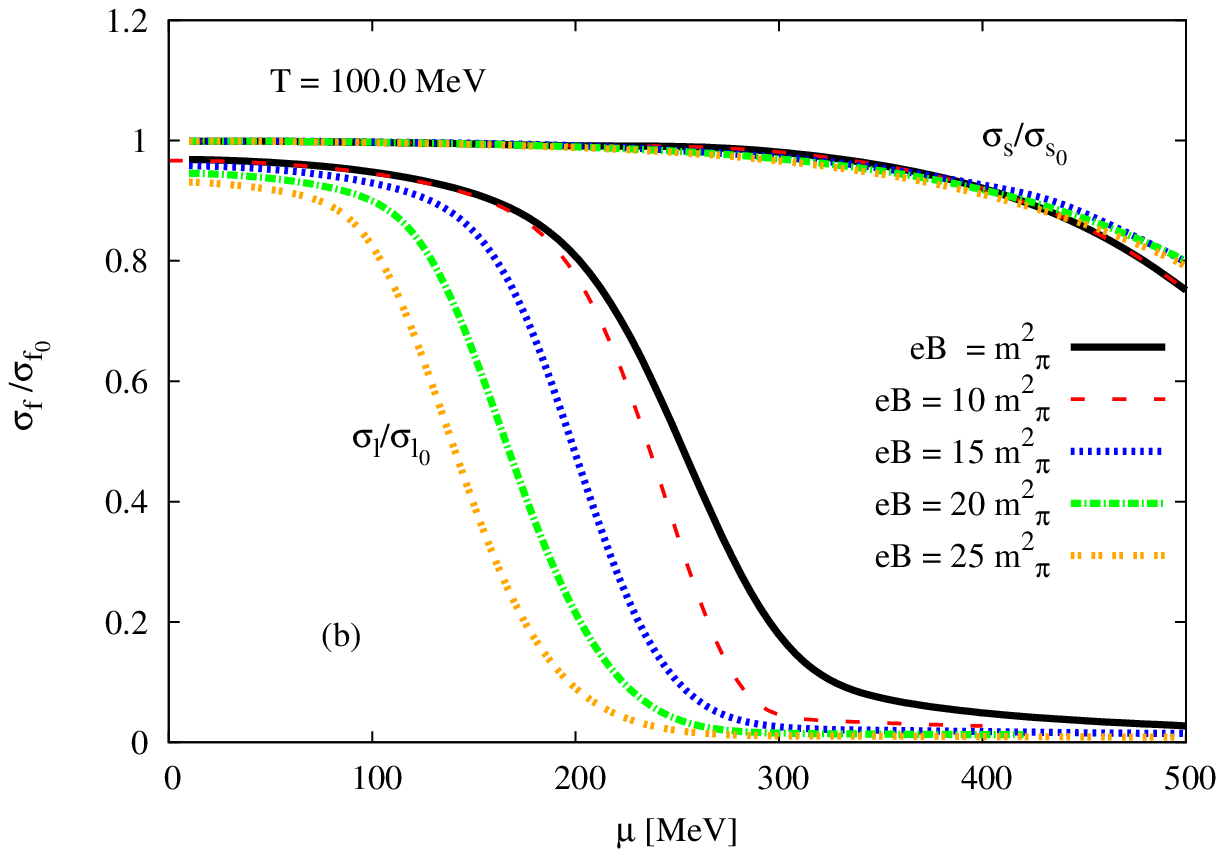}
\caption{$\sigma_l$ and $\sigma_s$ are given as functiona of the baryon chemical potential at different magnetic field strengths; $eB=1$ (solid curves) $10$ (dashed curves), $15$ (dotted curves), $20$ (dot-dashed curves), and $25\,m^2_{\pi}$ (double dotted curves). Left-hand panel (a) presents the results at $T=50~$MeV while right-hand panel (b) at $T=100~$MeV. \label{fig:sbtrcMu}
}}
\end{figure}

Fig. \ref{fig:sbtrcMu} presents $\sigma_l$ and $\sigma_s$ as functions of the baryon chemical potential at fixed temperatures; $T=50~$MeV (left-hand panel) and $T=100~$MeV (right-hand panel). Both chiral condensates are calculated in different magnetic field strengths; $eB=1$ (solid), $10$ (dashed), $15$ (dotted), $20$ (dot-dashed), and $25\,m^2_{\pi}~$ (double dotted curves). Although these $e B$ strengths are much larger than the one at NICA top energy ($e B \simeq 0.156\, m_{\pi}^2$), they reveal the tendency of $e B$ with the collision energies and the significant influence of the magnetic field strength can determined. 

We conclude that increasing temperature causes a rapid decrease in the chiral condensates, especially near the chiral phase-transition, Fig. \ref{fig:sbtrc1}. This is similar to the observation reported in a previous study without magnetic field \cite{Tawfik:2014gga}. At high density corresponding to NICA, there is a gap difference between light and strange chiral condensates, Fig. \ref{fig:sbtrcMu}. This can be understood because of the anomaly term in the mesonic part of the Lagrangian \cite{Tawfik:2014gga}. The anomaly term was proposed \cite{Schaefer:2008hk} to enhance the numerical estimation for the chiral condensates. Originally, this proposed term  models the axial U($1$)$_A$ anomaly of the QCD vacuum, which is likely broken by quantum effects. As the anomaly term explicitly appears in the strange quark condensate, we conclude that it considerably counts for its difference relative to the light quark condensate. At high temperature and density, it remains finite for the strange quark condensate while entirely vanishes for the light quark condensate. For the sake of completeness, we recall here that this anomaly term also appears in the meson masses. Its effects have been systematically analysed for sixteen meson states \cite{Tawfik:2014gga}. Also, here its effects at high density are remarkably large. From Fig. \ref{fig:eBdepence}, it is obvious that the magnetic field causes a considerable effect on the chiral phase-structure with varying temperature. Reducing the magnetic fields leads to an increase in the chiral critical-temperature, right-hand panel of Fig \ref{fig:eBdepence}. Furthermore, raising the magnetic field strength sharpens the QCD phase-structure and accelerates the formation of the metastable phase. The latter is to be understood due to early phase-transition, i.g. smaller critical temperature. 

Now, we analyse the influence of the magnetic field strength on the chiral phase-diagram. In Fig. \ref{fig:eBdepence}, the chiral temperature $T_c$ (left-hand panel) and baryon chemical potential (right-hand panel) are calculated in dependence on $e B$. $T_c$ is calculated according two criteria. The first one implements the second-order moment of the quark multiplicity, in which $T_c$ is determined at the peak of the normalized quark susceptibility ($\chi _q/T^2$). This criterion is frequently utilized in lattice QCD simulations. Depending on the order of the phase transition, the corresponding critical temperature can be determined. In cross-over, which is likely for light and strange quarks at low density, the phase-transition takes place in a range of temperatures. Thus, an average value can be assigned to the {\it quasi-}critical temperature. Our results are given by the solid curves. The second criterion defines $T_c$ as the intersect of the light-quark chiral condensate and the deconfinement order-parameter \cite{Tawfik:2014uka}. As illustrated in Figs. \ref{fig:sbtrc1} and  \ref{fig:sbtrcMu}, the chiral  {\it quasi-}critical temperature can be determined from $\sigma_l$ and $\sigma_y$, separately, which should become coincident with the deconfinement phase-transition. In PLSM, the latter is related to the temperature dependence of the order parameters $\phi$ and $\phi^*$. Both condensates seem to have cross-over phase transitions, where their values decrease from large values at low temperatures to low values at high temperatures. Accordingly, $T_c$ can be approximated as an average within the cross-over region. In order to increase the certainty of the estimated $T_c$, we utilize coincidence of both chiral condensates with $\phi$ and $\phi^*$. The corresponding results are depicted by dashed curves. The vertical bands mark $e B$ at $7.7$ GeV comparable to NICA ($\sim 0.156\, m_{\pi}^2$), RHIC beam energy scan ranging from $7.7$ to $200~$GeV  ($0.156 - 4.051\, m_\pi^2$) \cite{Kharzeev2015,Kharzeev2008} and $2760~$GeV at LHC ($\sim 10-15\, m_\pi^2$) calculated from PYTHIA with $30-40\%$ centrality and impact parameter $\sim 9~$fm.

We observe that $T_c$ decreases with increasing $e B$, i.e., inverse magnetic catalysis. A good agreement between dotted curve (second criterion) and the lattice results, especially at $0\leq\,eB\,\leq 0.2~$GeV$^2$, is obtained. At a wider range of $e B$; $0.13\leq\,eB\,\leq 0.55~$GeV$^2$, the solid curve (first criterion) matches well with the lattice calculations \cite{lattice:2014}. $T_c$ estimated from the first criterion apparently overestimates the lattice simulations, especially at small $e B$. The PLSM calculations refer to larger $T_c$ than the one deduced from the lattice QCD. On the contrary, the second criterion seems to underestimates $T_c$ (relative to the lattice) at large $e B$.

In right-hand panel (b), in a constant magnetic field such as the one marked by the vertical bands, we observe that, a larger $\mu$ can be reached at a lower temperature, i.e., $\mu$ decreases with increasing $e B$. In determining the critical baryon chemical potential at which the broken chiral symmetry should be restored, we implement the intersection between deconfinement phase-transition and the light-quark chiral condensate. At fixed temperatures, $T=50$ (solid curve) and $100~$MeV (dashed curve), the dependence of critical $\mu$ on $e B$ is illustrated in the right-hand panel of Fig. \ref{fig:eBdepence}. In determining the critical chemical potential, a second criterion, the intersection between deconfinement order-parameter and light-quark chiral condensate, can be implemented.

\subsection{Magnetic response and susceptibility}

As mentioned in previous sections, the nonzero magnetic field comes up with modifications to the dispersion relation so that the response of the QCD matter can be determined from the derivative of the free energy density $\mathcal{F}= - T/V \cdot \ln \mathcal{Z}$ with respect to the magnetic field itself
\begin{equation}
\mathcal{M}=- \frac{\partial \mathcal{F}}{\partial (eB)}, 
\end{equation}
where $e\neq 0$ is the elementary electric charge. The sign of the magnetization defines whether the QCD matter is {\it para}- or {\it dia}-magnetic, i.e. $M>0$ (bara-), or $M<0$ (dia-). Similar to solid-state physics, 
\begin{itemize}
\item if the QCD matter is {\it dia}-magnetic, then the color-charges are oriented oppositely to the magnetic field direction and an induced current shall be produced. The induced current spreads in form of small loops which - in tern - try to cancel out the magnetic effects, and 
\item if the QCD matter is  {\it para}-magnetic, then most color-charges shall be aligned towards the magnetic field direction. 
\end{itemize}

Recently, the magnetic susceptibility of the QCD matter has been reported \cite{susceptibility:2014}
\begin{equation}
\chi_B= - \frac{\partial^2 \mathcal{F}}{\partial (eB)^2}.
\end{equation} 
The magnetization is a dimensionless proportionality parameter indicating the degree of magnetization and depending on the magnetic permeability \cite{Hall}. Thus, it is apparent that the temperature weakens the magnetization and at relativistic energies (or vanishing chemical potential), the magnetization likely vanishes.

\begin{figure}[!htb]
\centering{
\includegraphics[width=7cm]{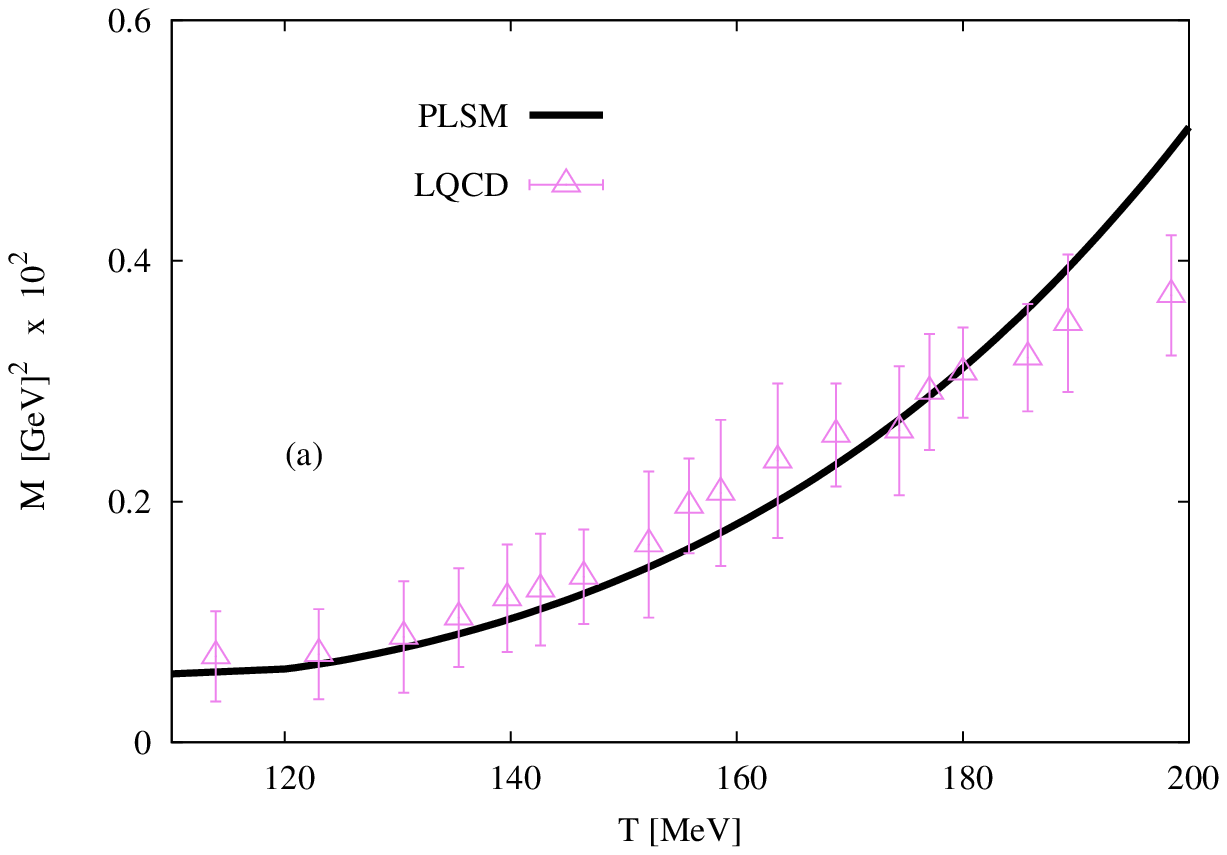}
\includegraphics[width=7cm]{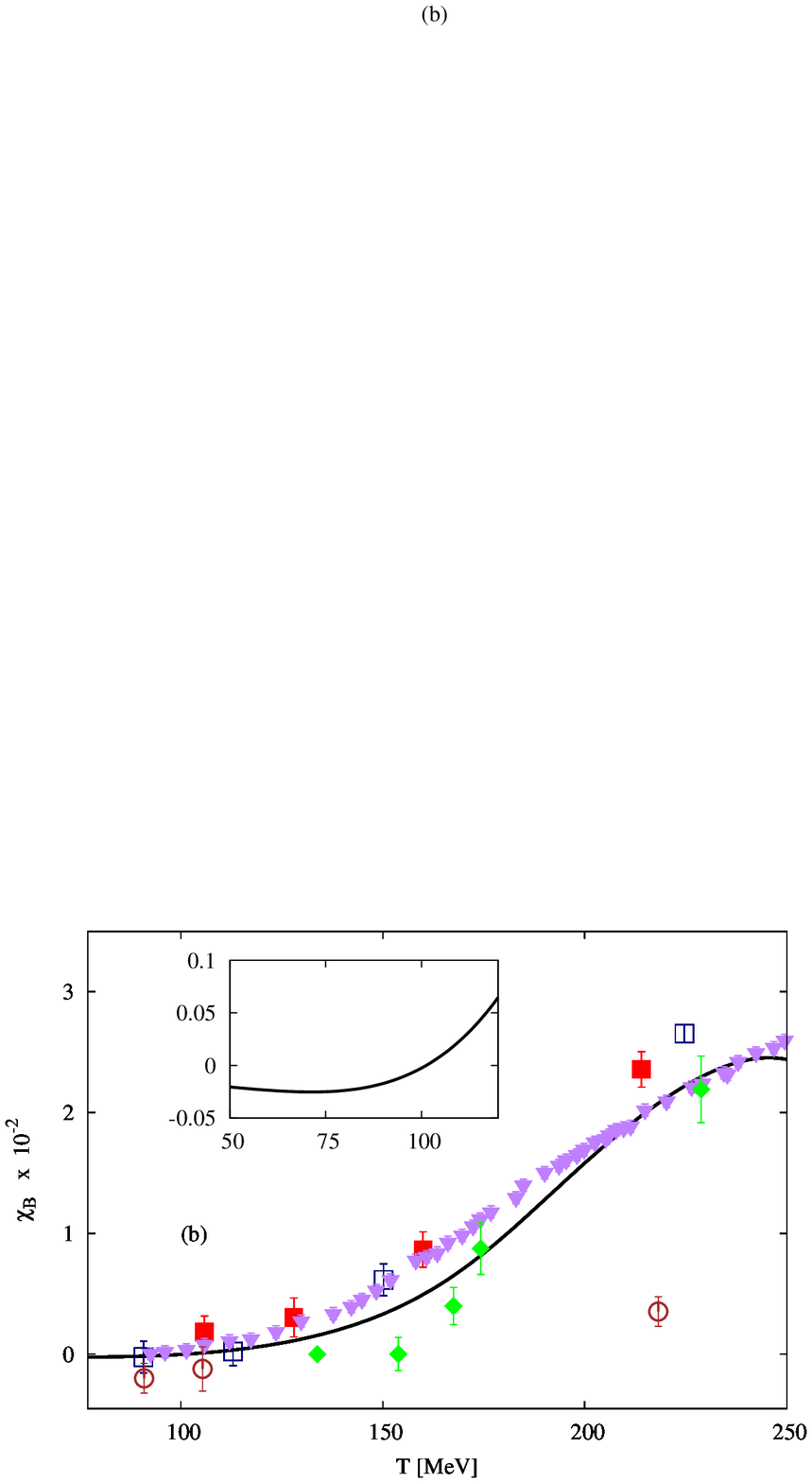}
\caption{The temperature dependence of the magnetization $\mathcal{M}$ [left-hand panel (a)], and of the magnetic susceptibility $\chi$ [right-hand panel (b)] are determined at $eB=0.2~$GeV$^2$ and a vanishing baryon chemical potential. The results are compared with different lattice simulations (symbols) \cite{lattice:2014}. \label{propes}
}}
\end{figure}

In left-hand panel of Fig. \ref{propes} (a), the magnetization of the QCD matter due to the effects of nonvanishing magnetic field $eB=0.2~$GeV$^2$  is studied as a function of temperature at a vanishing baryon chemical potential and compared with recent lattice calculations (open triangles with errorbars) \cite{lattice:2014}. Increasing the magnetization with the temperature obviously refers to positive magnetization, $M>0$, i.e., the paramagnetic property of the QCD matter becomes dominant. At temperatures smaller than the critical one, the calculations agree well with the lattice results. At temperatures larger than the critical one, we find that our PLSM calculations slightly overestimate the lattice results. In this limit, the degrees of freedom might not be sufficient to achieve a good agreement at very high temperature. Furthermore, such a discrepancy can be interpreted due to the PLSM applicability, which is determined up to the temperature range, where $\sigma_l$, $\sigma_s$, $\phi$ and $\phi^*$ remain finite. For the sake of completeness, we emphasize that the lattice results at $e B=0.2~$GeV$^2$ are obtained by using half-half method for three lattice temporal extensions; $N_\tau=6$ and $8$ and the continuum limit.

The right-hand panel of Fig. \ref{propes} (b) presents the magnetic susceptibility as a function of temperature at $eB=0.2~$GeV$^2$ and  a vanishing chemical potential. The results from PLSM are compared with various lattice simulations (symbols) in which different simulation methods and lattice configurations are applied. Features of PLSM and lattice QCD results can be summarized as follows.
\begin{itemize}
\item The PLSM free energy, Eq. (\ref{eq:plsmPOT}), consists of three parts. The first one is the gauge potential (pure meson). The second one represents the quark and antiquark potential, which includes fluctuations of quarks and antiquark. The third part takes into consideration the color and gluon interactions. This means that two types of contributions are not free of fluctuations, while the third one takes into consideration the gluon interactions.

\item At low temperatures, the slope of the magnetic susceptibility $\chi(T)$ becomes negative (inside-box in the right-hand panel). This signals a dia-magnetic property and apparently confirms the lattice results, as well. Such result has been also obtained from parton-hadron-string dynamics  \cite{Cassing:2014}. On the other hand, at high temperature, i.e., when the broken chiral symmetry is restored, a transition between dia- and para-magnetic properties is observed. Similarly, from non-interacting MIT bag model, the phase-transition from dia- to para-magnetism was observed \cite{MIT:2008}. QCD-like models with Polyakov potential agree well with the lattice calculations, especially at high temperatures \cite{Orlovsky:2014}.

\item The lattice simulations \cite{lattice:2014} implement half-half method in $24^3 \times 32$ lattice (closed triangle) and integral method in $28^3 \times 10$ lattice (open triangle). Thus, the continuum limit is likely achieved \cite{lattice:2013}. For $N_f=2+1$ degrees of freedom and by using HISQ/tree action with quark masses $m_l /m_s=0.05$ and $N_\tau=8$, the lattice results are represented by diamonds \cite{lattice:2013b}. The closed circles stand for isotropy lattice \cite{squeezing:2013}. 

\item The PLSM calculations seem to confirm the para-magnetic property of the QCD matter. At $100\leq\,T\,\leq 250~$MeV, the magnetic susceptibility steeply increases with approaching the deconfinement phase-transition \cite{Borsanyi:009,Borsanyi:010}.

\end{itemize}

\section{Conclusions} 

\begin{figure}[!htb]
\centering{
\includegraphics[width=7cm]{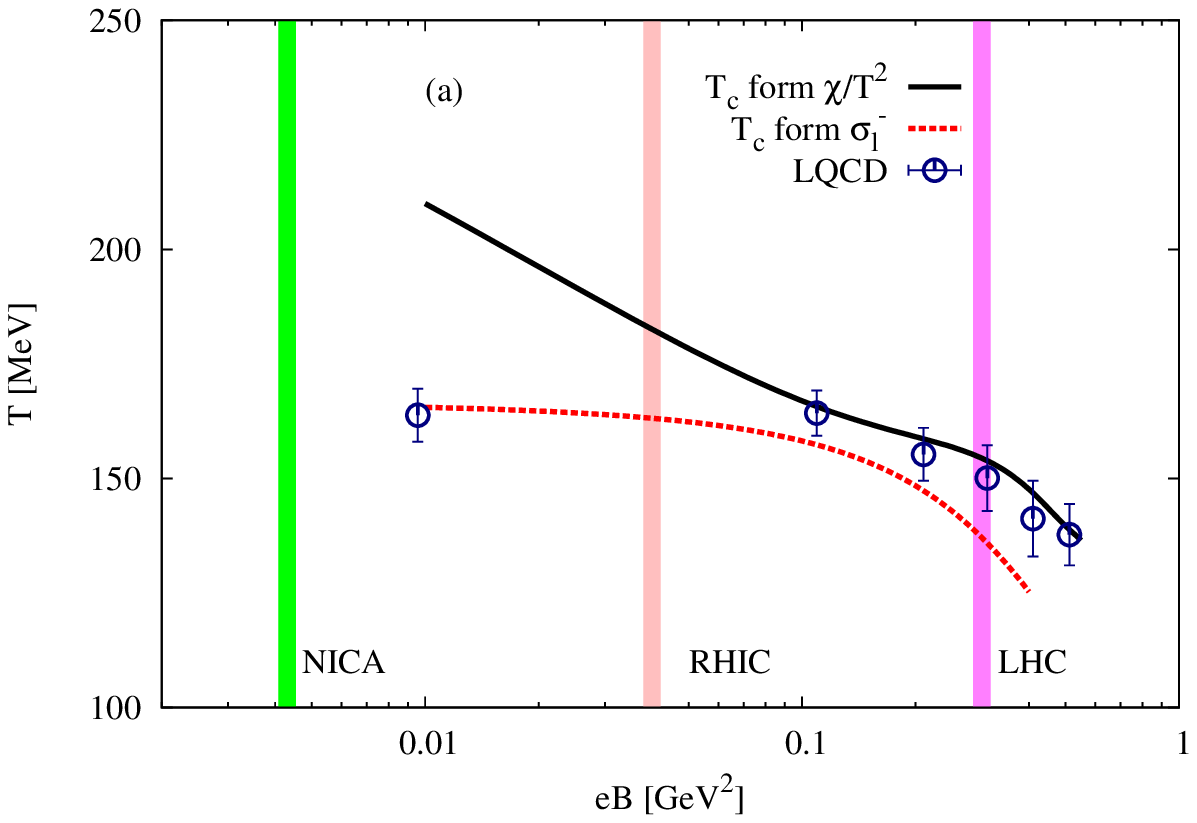}
\includegraphics[width=7cm]{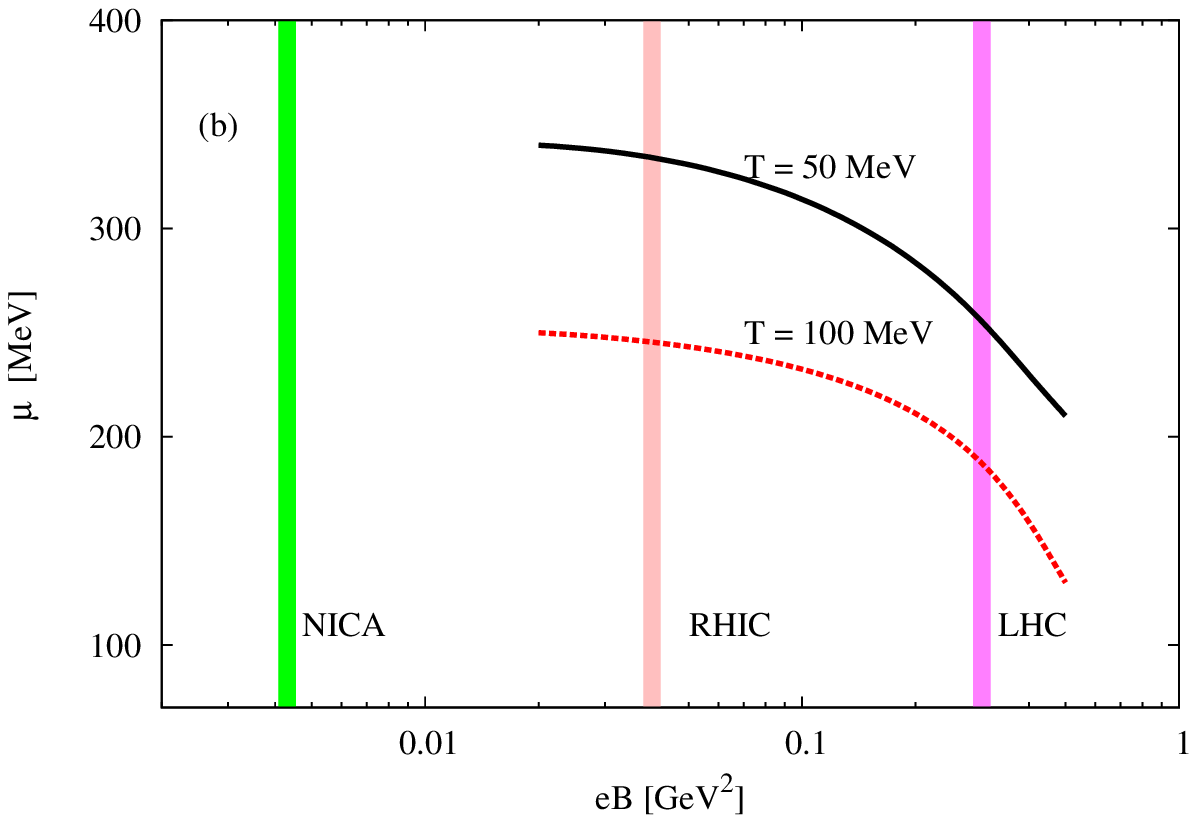}
\caption{Left-hand panel (a): the chiral critical temperature is depicted in dependence on $eB$ and compared with recent lattice calculations (symbols). The corresponding baryon chemical potential is given as a function of $eB$ [right-hand panel (b)]. The vertical bands refer to magnetic fields for NICA at $\sim 7.7~$GeV, for RHIC at $7.7-200~$GeV and for LHC at $2760~$GeV in heavy-ion collisions with $30-40\%$ centrality and impact parameter $b=9~$fm. RHIC band is apparently averaged.  \label{fig:eBdepence} 
}}
\end{figure}

Although the magnetic field strength in heavy-ion collisions at NICA energies is small relative to top RHIC and LHC energies, Fig. \ref{fig:eBdepence}, we observe that even such relative small $e B$ considerably decreases the critical temperatures with increasing magnetic field strength, i.e., inverse magnetic catalysis. At finite magnetic field, we find that the phase-transition seems to take place at higher critical temperatures than that of the restoration of the chiral symmetry breaking. The estimation of both types of critical temperatures is effectively achieved from the location (in temperature axis) of the second-order moment of the quark multiplicity, know as susceptibility and from the intersection of light and strange quark condensates with the deconfinement order parameters $\phi$ and $\phi^*$.

Both chiral condensates ($\sigma_l$ and $\sigma_s$), which can be characterized through thermal, dense and magnetic in-medium modifications of rho-meson, for instance \cite{TawfikINJP1}. We have presented results on their evolution with increasing temperature, chemical potential and magnetic field. We find that the chiral condensates rapidly decrease with increasing the baryon chemical potential, i.e., the chiral critical temperature decreases with the increase in the magnetic field strength. We conclude that the influences of the magnetic field are almost the same as that of the baryon chemical potentials, especially on decreasing the phase-transition temperature. Thus, the influence of the magnetic field at NICA energies ($e B \simeq 0.156\, m_{\pi}^2$) becomes stronger because of the high density. The susceptibilities and fluctuations of different particle yields are examples about possible experimental signatures for the deconfiment phase-transition and critical endpoint. The latter is conjectured to connect the cross-over of the QCD boundary at low density (chemical potential) and the first-order phase transition at high density \cite{PLSMINJP1}. In light of this, we recommend systematic measurements for the phase-transition characterizing signals such as the fluctuations and the dynamical correlations in various interacting systems and centralities. These shall reveal a complete description for the QCD phase-structure and for the chemical freezeout at high density.

The magnetization and magnetic susceptibility almost monotonically increase with increasing temperature. At low temperatures, the magnetic susceptibility has negative slope, i.e., the QCD matter possesses dia-magnetic property. At high temperature, i.e., when the broken chiral symmetry is restored, the QCD matter becomes paramagnetic. In both temperature regimes, the PLSM calculations and the recent lattice calculations are in good agreement. In order to justify this agreement, we have shortly reviewed the configurations and the characteristics of the lattice QCD calculations.

Furthermore, we conclude that increasing critical temperature (as a result of decreasing magnetic field) is accompanied by a considerable decrease in the baryon chemical potential. We have analysed how at a fixed temperature the increase in $e B$ decreases the chiral critical temperature, which - in turn - decreases further with increasing the temperature. Nearly the same effects has the baryon chemical potential.

\begin{thebibliography}{99}

\bibitem{prephiral2} Z-T Liang and X-N Wang 
{\it Phys. Rev. Lett.} {\bf 94} 102301 (2005)

\bibitem{prephiral3} F Becattini, F Piccinini and J Rizzo 
{\it Phys. Rev. } {\bf C77} 024906 (2008)

\bibitem{prephiral1} D E Kharzeev, L D McLerran and H J Warringa 
{\it Nucl. Phys. } {\bf A803} 227 (2008)

\bibitem{fraga} E S Fraga and A J Mizher 
{\it Phys. Rev. } {\bf D78}, 025016 (2008)

\bibitem{Tawfik:quasi} A Tawfik and N Magdy 
{\it J. Phys. } {\bf G42} 015004 (2015)  

\bibitem{Tawfik:2014uka} A Tawfik, N Magdy and A Diab 
{\it Phys. Rev. } {\bf C89} 055210 (2014)

\bibitem{Tawfik:2014gga} Abdel Nasser Tawfik and  Abdel Magied Diab 
{\it Phys. Rev.  } {\bf C91} 015204 (2015) 

\bibitem{Tawfik:2014hwa} Abdel Nasser Tawfik and Niseem Magdy 
{\it Phys. Rev.  } {\bf C90} 015204 (2014)

\bibitem{Shovkovy2013} I A Shovkovy 
{\it Lect. Notes Phys.} {\bf 871} 13 (2013)

\bibitem{Schaefer:2008hk} B-J Schaefer and M Wagner 
{\it Phys. Rev. }~{\bf D79} 014018 (2009)


\bibitem{Kharzeev2015}  D E Kharzeev, J Liao, S A Voloshin, and G Wang 
{\it Prog. Part. Nucl. Phys.} {\bf 88} 1 (2016) 

\bibitem{Kharzeev2008} K Fukushima, D E Kharzeev and H J Warringa 
{\it Phys. Rev. Lett.} {\bf 104} 212001 (2010)


\bibitem{lattice:2014} G Bali, F Bruckmann, G Endrodi, S Katz and A Schaefer 
{\it J. High Energy Phys.}  {\bf 1408} 177 (2014)

\bibitem{susceptibility:2014} K Kamikado and T Kanazawa 
{\it J. High Energy Phys.} {\bf 1501} 129 (2015).

\bibitem{Hall} J R Hook and H E Hall {\it Solid state physics} (Chichester : Wiley) {\bf Vol 1} (1995)

\bibitem{Cassing:2014} T Steinert and W Cassing 
{\it Phys. Rev. } {\bf C89} 035203 (2014)

\bibitem{MIT:2008} N Agasian and S Fedorov 
{\it Phys. Lett. } {\bf B663} 445 (2008)

\bibitem{Orlovsky:2014} V Orlovsky and Y A Simonov 
{\it Int. J. Mod. Phys. } {\bf A30} 1550060 (2015)

\bibitem{lattice:2013} C Bonati, M D'Elia, M Mariti, F Negro and F Sanfilippo 
{\it Phys. Rev. } {\bf D89} 054506 (2014)

\bibitem{lattice:2013b} L Levkova and C DeTar 
{\it Phys. Rev. Lett.} {\bf 112} 012002 (2014)

\bibitem{squeezing:2013} G Bali, F Bruckmann, G Endrodi and A Schafer 
{\it Phys. Rev. Lett.} {\bf 112} 042301 (2014)

\bibitem{Borsanyi:009} Y Aoki, S Borsanyi, S Durr, Z Fodor, S D Katz, S Krieg and K K Szabo 
{\it J. High Energy Phys.} {\bf 0906} 088 (2009)

\bibitem{Borsanyi:010} S Borsanyi, G Endrodi, Z Fodor, A Jakovac, S D Katz, S Krieg, C Ratti, and K K Szabo 
{\it J. High Energy Phys.} {\bf 1011} 077 (2010)

\bibitem{TawfikINJP1} A Tawfik {\it Indian J. Phys.} {\bf 85} 755 (2011)

\bibitem{PLSMINJP1}  U S Gupta, R K Mohapatra, A M Srivastava and V K Tiwari {\it Indian J. Phys.} {\bf 85} 115 (2011)

\end{thebibliography}
\end{document}